\documentclass[aps,prb,twocolumn,superscriptaddress,showpacs]{revtex4}
\usepackage{graphicx}% Include figure files

\begin{document}

\title{Dynamical study on polaron formation in a metal/polymer/metal structure}
\author{C.Q. Wu}
\affiliation{Research Center for Theoretical Physics, Fudan
University, Shanghai 200433, China} \affiliation{Institute of
Materials Structure Science, KEK, Tsukuba, Ibaraki, 305-0801,
Japan}
\author{Y. Qiu }
\affiliation{Research Center for Theoretical Physics, Fudan
University, Shanghai 200433, China}
\author{K. Nasu}
\affiliation{Institute of Materials Structure Science, KEK,
Tsukuba, Ibaraki, 305-0801, Japan}

\date{\today}

\begin{abstract}
By considering a metal/polymer/metal structure within a tight-binding
one-dimensional model, we have investigated the polaron formation in the
presence of an electric field. When a sufficient voltage bias is applied to
one of the metal electrodes, an electron is injected into the polymer chain,
then a self-trapped polaron is formed at a few hundreds of femtoseconds
while it moves slowly under a weak electric field (not larger than $%
1.0\times 10^4$ V/cm). At an electric field between $1.0\times 10^4$ V/cm and $%
8.0\times 10^4$ V/cm, the polaron is still formed, since the
injected electron is bounded between the interface barriers for
quite a long time. It is shown that the electric field applied at
the polymer chain reduces effectively the potential barrier in the
metal/polymer interface.
\end{abstract}


\maketitle

\section{introduction}

Recent years, organic electronic devices (OEDs), e.g.,
light-emitting diodes (LEDs) and field-effect transistors (FETs),
are attracting considerable interest because they have processing
and performance advantages for low-cost and large-area
applications.\cite{campbell} In these devices, organic polymers
are used as the light-emitting and charge-transporting layers, in
which the electron and/or hole are injected from the metal
electrodes and transported under the influence of an external
electric field. Due to the strong electron-lattice interactions,
it is well known that additional electrons or holes in conjugated
polymers will induce self-localized excitations, such as
solitons\cite{ssh} (only in \textit{trans}-polyacetylene) and
polarons.\cite{brazovskii} As a result, it has been generally
accepted that the charge carriers in conjugated polymers are these
excitations including both charge and lattice
distortion.\cite{heeger} The formation and transport of such
charge carriers are believed to be of fundamental importance for
these OEDs.

There have been extensive studies on soliton and polaron dynamics
in conjugated polymers\cite{su,ono,rakhmanova,johansson} under the
influence of external electric fields. It is shown that solitons
as well as polarons keep their shape while moving along a chain.
Solitons are shown to have a maximum velocity $2.7 v_s$, where
$v_s$ is the sound velocity.\cite{bishop,ono} The situation will
be different for polarons, which has been shown to be not created
in electric fields over $6\times 10^4$ V/cm due to the charge
moving faster and not allowing the distortion to
occur.\cite{rakhmanova} A recent study by Johansson and
Stafstr\"{o}m\cite{johansson} deals with the polaron migration
between neighboring polymer chains. The numerical results show
that the polaron becomes totally delocalized, either before or
after the chain jump for the electric field over $3\times 10^5$
V/cm. There are also studies on the charge transport through
metal/polymer interfaces, e.g., the resistance of organic
molecular wires attached to metallic surface\cite {samanta},
Schottky energy barriers in metal/organic\cite{campbell2}, and the
dynamics of charge transport in a short oligomer sandwiched
between two metal contacts.\cite{yu}

In this paper, we present our results of a non-adiabatic dynamical
study on the polaron formation starting from charge injection in a
metal/polymer/metal structure in the presence of an external
electric field. We would like to focus on charge and its induced
lattice distortion in the polymer chain while the metal electrodes
at the two ends serves as the source and drain for the extra
charges within a tight-binding one-dimensional model. The electron
wavefunction is described by the time-dependent Schr\"{o}dinger
equation, in which the transition between instantaneous
eigenstates are allowed, while the polymer lattice is treated
classically by a Newtonian equation of motion.\cite{ono}

The paper is organized as follows. In the following section, we
present a tight-binding one-dimensional model for the
metal/polymer/metal structure and describe the dynamical evolution
method. Our results are presented in Sec. III and the summary of
this paper is given in Sec. IV.

\section{model and method}

We consider a one-dimensional metal/polymer/metal structure that
contains a polymer chain as well as two metal electrodes attached
at its two ends. The Hamiltonian consists of three parts,
\begin{equation}
H=H_e+H_{latt}+H_{ext}.
\end{equation}
The electronic part is
\begin{equation}
H_e=-\sum_n t_n(c_n^\dagger c_{n+1}+h.c.),
\end{equation}
$t_n$ being the hopping integral between site $n$ and $n+1$, that
is, $t_n=t_0$ in the two metal electrodes, $t_n=t_0-\alpha
(u_{n+1}-u_{n})$, where $u_n$ is the monomer displacement of site
$n$ and $\alpha$ describes the electron-lattice coupling between
neighboring sites in the polymer chain, as the
Su-Schrieffer-Heeger (SSH) model\cite{ssh}, and $t_n=t_1$ for the
coupling between sites connecting the polymer chain and the metal
electrodes. The polymer lattice is described by
\begin{equation}
H_{latt}=\frac{K}{2}\sum_n
(u_{n+1}-u_{n})^2+\frac{M}{2}\sum_n\dot{u}_n^2,
\end{equation}
where $K$ is the elastic constant and $M$ the mass of a CH
group.\cite{ssh} The contribution from the external field is
\begin{equation}
H_{ext}=\sum_{n}V_n(t)(c_n^\dagger c_n-1),
\end{equation}
$V_n(t)$ being site-energies due to the applied voltage bias and
electric field. At the left metal electrode a voltage bias is
applied for the charge injection, $V_n(t)=V(t)$. At the polymer
chain an uniform electric field $E(t)$ is applied,
$V_n(t)=|e|E(t)[(n-n_0)a+u_n]$ with $e$ being the electron charge,
$n_0$ the first site of the polymer chain, and $a$ the lattce
constant. At the right metal electrode, a field-free area, the
site-energies $V_n(t)$ are chosen as $V_{n_1}(t)$, $n_1$ is the
last site of the polymer chain. The spin index in the electron
operators is omitted since the electron interaction will not be
considered here. A similar static model is used very recently to
investigate the ground-state properties of ferromagnetic
metal/conjugated polymer interfaces.\cite{xie}

We consider a finite system containing a 200-monomer polymer chain
and two 100-site metal electrodes attached at the two ends. The
model parameters are those generally chosen for
\textit{trans}-polyacetylene:\cite{ssh,heeger} $t_{0}=2.5$eV,
$\alpha =4.1$eV$/\AA$, $K=21$eV$/\AA^2$,
$M=1349.14$eVfs$^{2}/\AA^2$, and $a=1.22\AA$. The coupling between
the polymer and the metal is set to $t_{1}=1.0$eV. Before we go
further for the dynamical evolution, we determine the static
structure of energy levels in a constant external field
($V(t)\equiv V_0$ and $E(t)\equiv E_0$) in this section.

The total energy is obtained by the expectation value of the
Hamiltonian (1) at the half-filled ground state $|g\rangle$,
\begin{equation}
E_t=\langle
g|H_e+H_{ext}|g\rangle+\frac{K}{2}\sum_n(u_{n+1}-u_n)^2.
\end{equation}
The electronic states are determined by the electronic part of the
Hamiltonian (1) and the lattice configuration of the polymer
$\{u_n\}$ is determined by the minimization of the total energy in
the above expression,
\begin{equation}
u_{n+1}-u_n=-\frac{\alpha}{K}(\rho_{n,n+1}+\rho_{n+1,n})
+\frac{|e|E_0}{K}\rho_{n,n}+\lambda,
\end{equation}
where $\lambda$ is a Lagrangian multiplier to guarantee the
polymer chain length unchanged, i.e., $\sum_n(u_{n+1}-u_n)=0$.
$\rho_{n,n'}$ is the element of density matrix, which will be
given below.

\begin{figure}
\includegraphics[angle=270,scale=0.36]{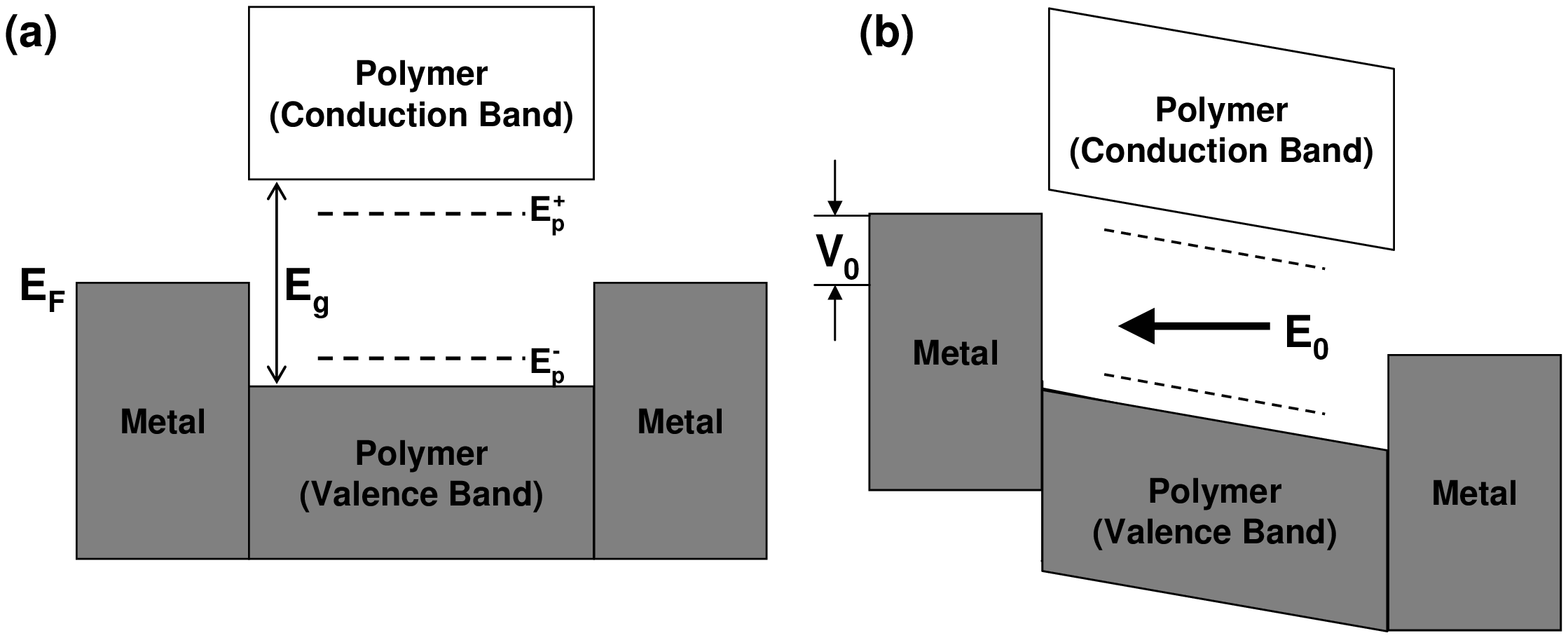}
\caption{Schematic energy level diagrams of the
metal/polymer/metal structure in the absence (a) and presence (b)
of external fields. $E_g$ is the band gap. The two dashed lines
indicate the localized levels if a polaron exists in the polymer
chain. $V_0$ is the voltage bias applied at the left metal
electrode and $E_0$ is the electric field applied at the polymer
chain. \label{fig1}}
\end{figure}

Due to the coupling between the polymer and the metal electrodes,
the electronic states in the polymer and the metal electrodes are
mixed. By calculating the wave function possibilities
$\eta_\mu^{(\kappa)} (\equiv\sum_{n\in\kappa}|\phi_\mu(n)|^2)$ for
each state $\mu$, where $\phi_\mu(n)$ is the wave function at site
$n$ and $\kappa$ is a set of sites (the left metal electrode, the
polymer chain, or the right metal electrode) in the system, we can
find that each electronic state can be clearly distinguished to be
in the polymer or the metal electrodes. It indicates that the
coupling $(t_1/t_0)$ in the metal/polymer interfaces is small.

In Fig.~1, we show the schematic energy level diagrams. In the
absence of external fields, the polymer chain is dimerized for a
half-filled system, the energy band, as in a single polymer chain,
has a gap of about $E_g=1.32$eV, and the metal electrodes have
quasi-continuum bands with the Fermi level almost at the middle of
the gap. The dashed lines indicates the localized levels if a
polaron exists in the polymer chain. $E_p^\pm\approx\pm 0.45$eV,
indicates the binding energy of a polaron to be about $0.21$eV. It
has been known that the energy gap of MEH-PPV is
$2.4$eV\cite{hag}, a much larger value than the energy gap $E_g$
we obtained above, but the polaron binding energy is similar to
that of MEH-PPV.

In the presence of a constant external field (the voltage bias
$V_0$ and the electric field $E_0$), the energy levels will be
changed as shown in Fig.~1(b). The voltage bias $V_0$ applied at
the left metal electrode raises its energy levels by a value of
about $V_0$ while the energy levels in other parts are almost
unchanged at $E_0=0$. It has been shown\cite{qiu} that when the
highest occupied electronic state of the left metal electrode is
close to the bottom of the conduction band due to the applied
voltage bias ($V_0 > 0.56$eV), the system will be unstable and
charges start to be injected into the polymer through the
metal/polymer interface. It will be seen that the electric field
$E_0$ applied at the polymer chain favors the charge injection as
a result of the incline of the site-energy at polymer chain.

The voltage bias $V_0$ could also serve as the metal work
function. At $V_0=0$, the energy difference between the electron
polaron level in the polymer chain and the highest occupied level
of the metal electrode is about 0.45eV. For the contact with
MEH-PPV and aluminum, the energy difference is about 1.4eV, and
the value is about 0 for that with MEH-PPV and calcium.\cite{hag}
So the value $V_0=0.57$eV, we take in the following calculation,
could be thought to correspond to that for the Ca/MEH-PPV contact
with the voltage bias about 0.2eV applied at the metal calcium.

Now, we describe the non-adiabatic dynamical method that has been
used for the dynamics of soliton\cite{ono} and
polaron\cite{rakhmanova,johansson} in an electron-lattice
interacting system. The evolution of the electron wavefunctions
depends on the time-dependent Schr\"{o}dinger equation
\begin{equation}
i\hbar \dot{\phi}_{n,\mu}(t)=-t_{n-1}\phi _{n-1,\mu}(t)+V_n(t)\phi
_{n,\mu}(t)-t_n\phi_{n+1,\mu}(t),
\end{equation}
where the site index $n$ runs over the whole chain. The lattice
displacements are determined classically by the following
Newtonian equations of motion
\begin{eqnarray}
M\ddot{u}_{n}(t)&=&K[u_{n+1}(t)+u_{n-1}(t)-2u_{n}(t)]\nonumber\\
&&+2\alpha[\rho_{n,n+1}(t)-\rho_{n-1,n}(t)]\nonumber\\
&&+|e|E(t)[\rho_{n,n}(t)-1],
\end{eqnarray}
where $n$ runs only in the polymer sites. $\rho_{n,n'}$ is the
element of the density matrix defined as
\begin{equation}
\rho_{n,n'}(t)=\sum_{\mu}\phi_{n,\mu }^*(t)f_{\mu }\phi_{n',\mu
}(t),
\end{equation}
where $f_{\mu}$ is the time-independent distribution function
determined by initial occupation (being 0, 1, or 2). The coupled
differential equations (7) and (8) can be solved numerically by
use of the same technique in Ref.\cite{ono,johansson}. The time
step is chosen to be as small as $0.1$fs to avoid numerical
errors. A small damping is also introduced to smear out the
lattice vibrations, and the effect of the vibrations will be
elaborated elsewhere.

In the real calculation, we choose the external field to be turned
on smoothly, that is, we let $V(t)=V_0\exp[-(t-t_c)^2/t_w^2]$ for
$0 < t < t_c$ and $V(t)=V_0$ for $t\geq t_c$ with $t_c$ being a
smooth turn-on period and $t_w$ the width. As the voltage bias
$V(t)$, the applied electric field $E(t)$ is also turned on
smoothly, i.e., $E(t)=E_0\exp[-(t-t_c)^2/t_w^2]$ for $0 < t < t_c$
and $E(t)=E_0$ for $t\geq t_c$ with the same $t_c$ and $t_w$. In
calculations, we take $t_{c}=30fs$, $t_{w}=25fs$, and various
values of voltage bias $V_0$ and electric field $E_0$.

\section{results}

In this section, we present our results on the polaron formation
starting from the charge injection from the metal electrode in the
presence of an external electric field. For that, we put one extra
electron on the lowest unoccupied state of the left metal
electrode while the voltage bias and electric field are turned on.
In the following, we will focus on the evolution of the charge
distribution $\rho_n$ [$\equiv \rho_{n,n}-1$] and the staggered
lattice configuration $y_n$ [$\equiv (-1)^n(u_n-u_{n-1})$]. At the
initial state $(t=0)$, the polymer chain has a dimerized lattice
of $y_n\approx 0.08\AA$ but a few bonds at its two ends and
$\rho_n=0$ for all sites in the polymer chain.

First of all, we shows the charge distribution $\rho_n$ and the
staggered lattice configuration $y_n$ (in unit of $\AA$) of the
polymer chain at a few typical times under a weak electric field
$E_0=5\times 10^3$V/cm in Fig.~2. The voltage bias applied at the
left metal electrode is taken as $V_0=0.57$eV, which is just above
the minimum value of the bias for the charge injection in the
absence of the electric field.\cite{qiu} For a smaller value of
$V_0$, the charge will not be injected into the polymer since the
energy of the highest occupied electronic level at the left metal
electrode ($E_{HO}$) is much lower than the bottom of the
conduction band of the polymer ($E_g/2$). For a larger value of
$V_0$, more electrons will be transferred into the polymer since
more occupied levels of the left metal electrode will be higher
than $E_g/2$.

\begin{figure}
\includegraphics[angle=270,scale=0.32]{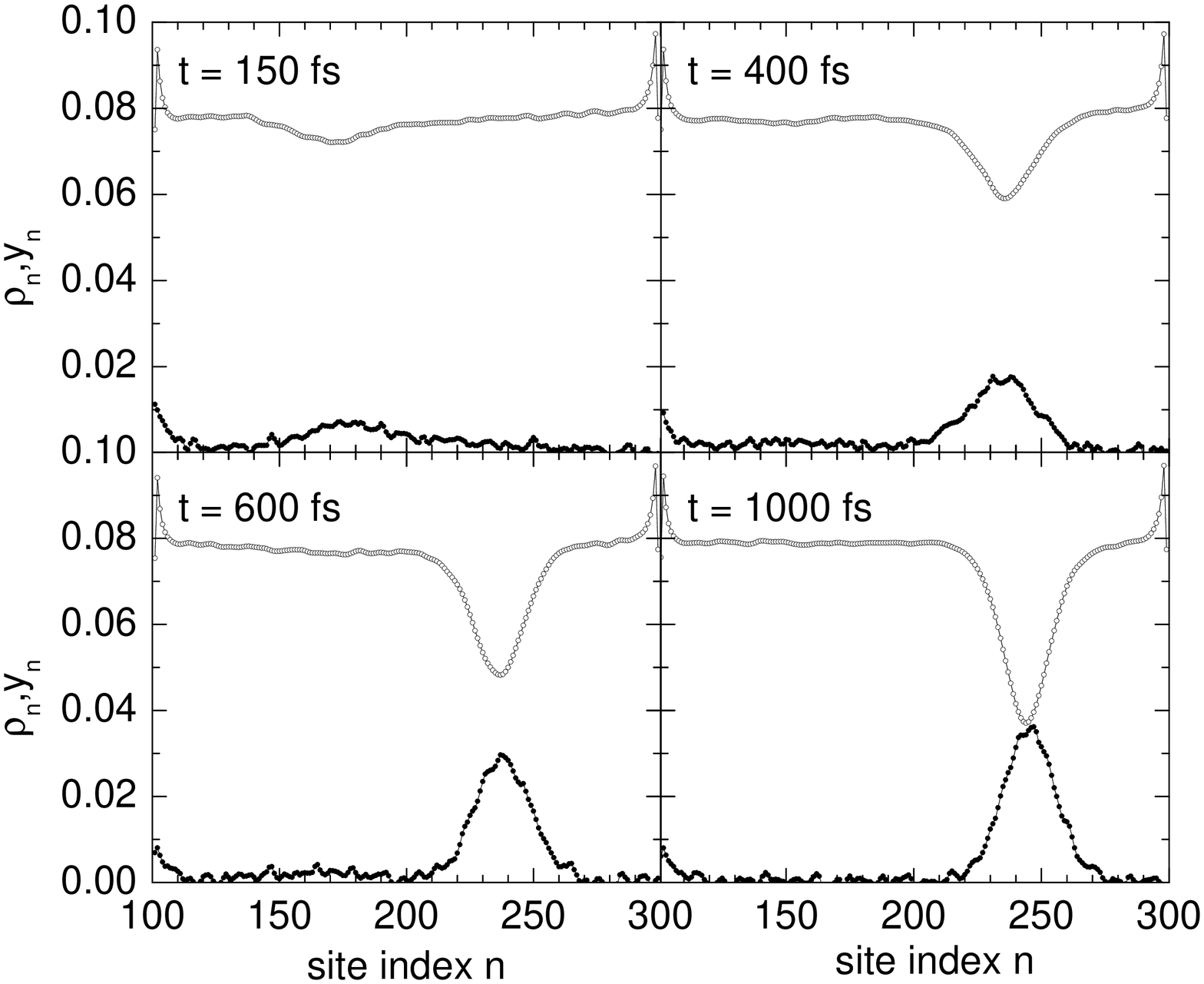}
\caption{The charge distribution $\rho_n$ (solid circles) and the
staggered lattice configuration $y_n$ (open circles) of the
polymer chain at different times under a weak electric field
$E_0=5\times 10^3$V/cm. $V_0=0.57$eV. \label{fig2}}
\end{figure}

As the voltage bias applied at the left metal electrode increases
to $V_0=0.57$eV smoothly, $E_{HO}$ goes up toward the bottom of
the conduction band of polymer. At $t=30$fs, $E_{HO}$ reaches its
maximum value $0.65$eV, which is very close to the bottom of the
conduction band. As a consequence, the coupling between energy
levels mainly at the left metal electrode and at the polymer
becomes stronger and stronger, the extra electron initially on the
left metal electrode begins to be injected into the polymer
gradually. From the Fig.~3, which shows the evolution of the total
charges in the polymer chain and the metal electrodes, we can see
that the charge injection takes a few hundreds of femtoseconds,
which seems to be insensitive to the strength of electric fields.

Furthermore, from Fig.~3(a), one can find that the charge does not
enter the right metal electrode, which shows the electric field is
too weak to overcome the barrier at the right polymer/metal
interface, though we have made the calculation up to a few ms. The
situation will be different for the electric field $E_0>1.0\times
10^4$V/cm, see Fig.~3(b), which is strong enough for the charge to
overcome the interface barrier and finally enter the right metal
electrode in a few thousands fs.

\begin{figure}
\includegraphics[angle=0,scale=0.35]{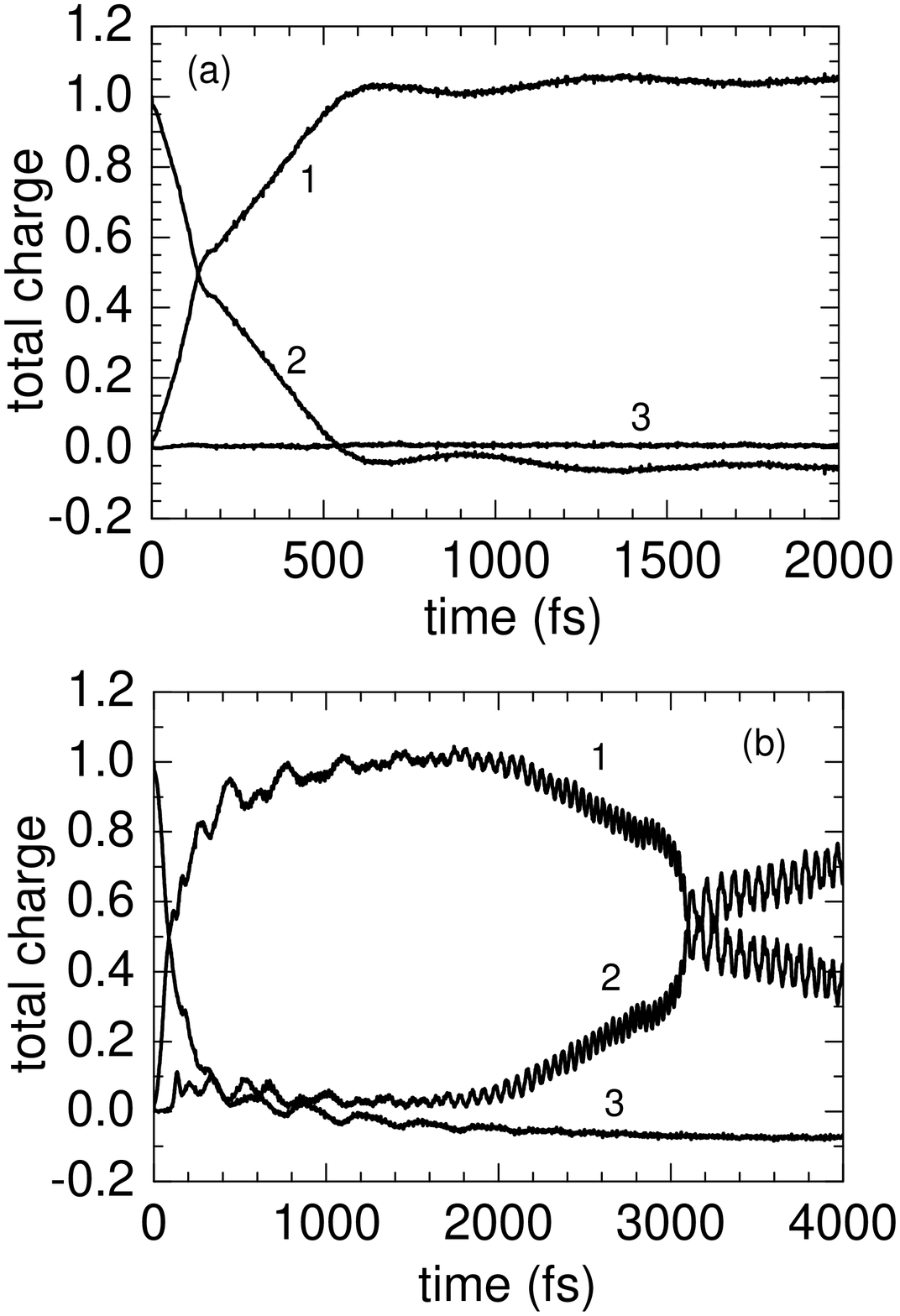}
\caption{Evolution of the total charges in the polymer chain (line
1) and the metal segments (line 2 and 3) at the electric field
$E_0=5\times 10^3$V/cm (a) and $E_0=3.0\times 10^4$V/cm (b).
$V_0=0.57$eV.\label{fig3}}
\end{figure}

In contrast with the case of $E_0=0$,\cite{qiu} the polaron is
formed not at the left side of the polymer but at the right side,
as shown in Fig.~2, due to the action of the electric field. It
will be seen more clearly from Fig.~4, which shows the evolution
of the charge center $x_c$ in the polymer, which is defined as
\begin{equation}
x_c=\sum_nn\rho_n/\sum_n\rho_n,
\end{equation}
where the summation is over the polymer sites. It can be seen that
there are two regions, one is the charge injection at $t=0 -
200$fs, and the other is the polaron formation at $t=200-500$fs.
In the first region, the charge is injected into the extended
states of the polymer and moves under the electric field. Since
the localized lattice distortion has not yet formed in this
region, the injected electron behaves as a free conduction
electron. As a comparison, we also show the curve for $E_0=0$ in
Fig.~4, where the charge moves away from the interface just due to
the energetic favorite. At a weak electric field, such as
$E_0=5.0\times 10^3$V/cm, the charge center moves gradually to the
middle of the polymer chain at $t=200$fs, which corresponds with
the charge being almost uniform in the polymer. However, at an
electric field with the strength $E_0$ between $1.0\times
10^4$V/cm and $8.0\times 10^4$V/cm, the charge will be spread
quickly and then bounced back at the right polymer/metal
interface, which is also the reason that the polaron will be
formed at a position closer to the left interface under a stronger
electric field. In the second region $t=200-500$fs, the lattice
distortion makes the polaron form with the charge being trapped
inside while it moves toward to the right interface under the
electric field.

\begin{figure}
\includegraphics[angle=270,scale=0.32]{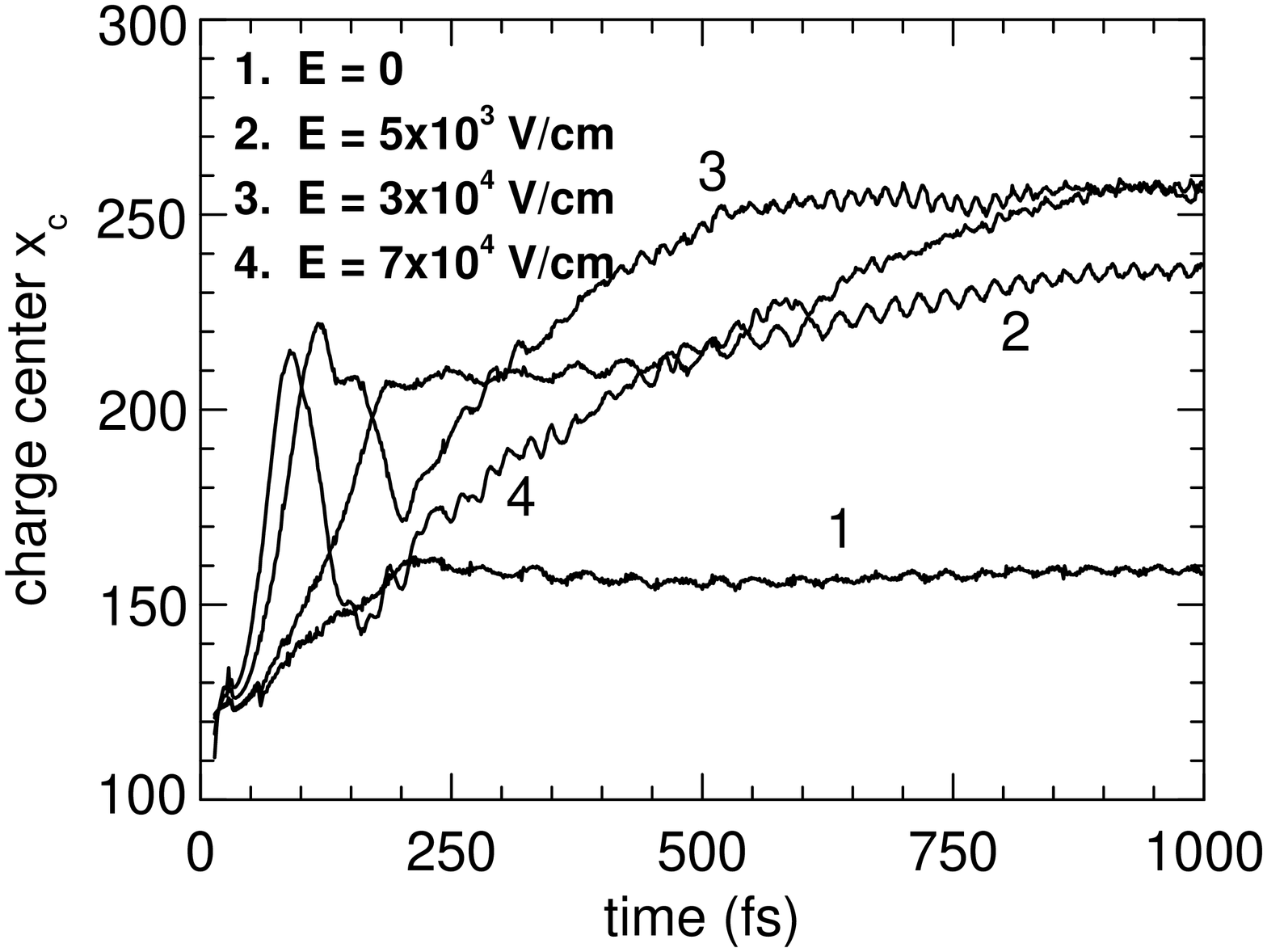}
\caption{Evolution of the charge center $x_c$ in the polymer chain
for various electric fields. $V_0=0.57$eV.\label{fig4}}
\end{figure}

What we have found also implies that the polaron will not form in
electric fields $\geq 10^4$V/cm if the charge does not be bounded
by the metal/polymer interfaces. This critical strength of applied
electric fields is smaller than that obtained by Rakhmanova and
Conwell\cite{rakhmanova} who gave the value of $6\times 10^4$V/cm
for the polaron formation but with the initial state of the
electron being injected into the polymer chain with the lowest
possible energy.

\begin{figure}
\includegraphics[angle=270,scale=0.32]{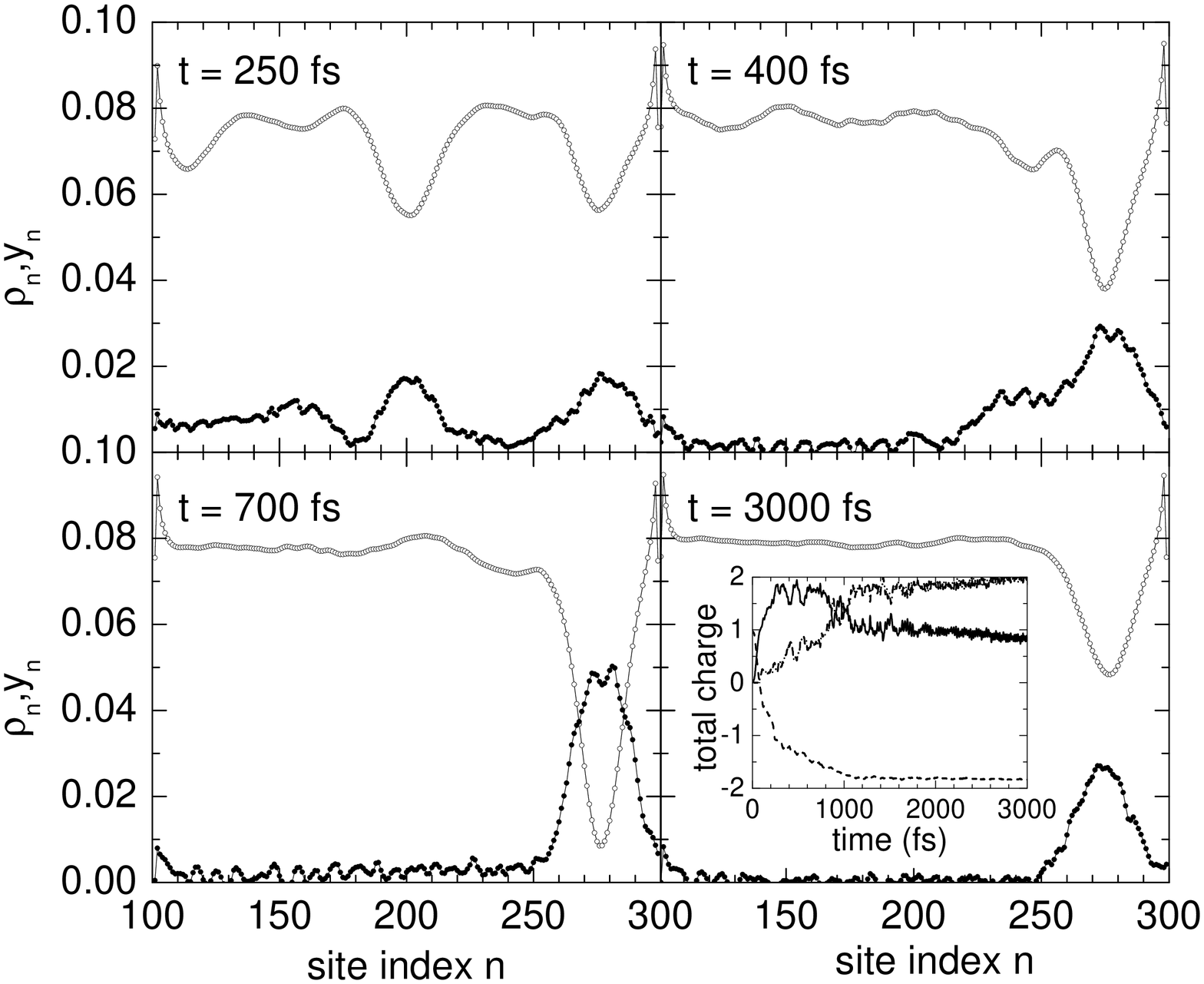}
\caption{Same as in Fig.~2 but with $E_0=1.0\times 10^5$V/cm and
$V_0=0.57$eV. The inset shows the total charges in the polymer
chain (solid line) and the metal electrodes (dash and dot-dash
lines). \label{fig5}}
\end{figure}

In the case of the electric fields $E_0\ge 8.0\times 10^4$V/cm,
there will be more electrons being injected into the polymer as a
result of the incline of the site-energy at polymer chain. In
Fig.~5, we show the evolution of the polaron formation for
$E_0=1.0\times 10^5$V/cm. It can be seen from the inset of Fig.~5
that the total charge at the polymer has reached $2e$ at
$t=250$fs, when the lattice displays a few distorted wells at
random locations. When the charge continues to be injected into
the polymer at the period $t=250-1000$fs, the lattice forms a
single deep well close to the right interface and the charge also
starts to enter the right metal electrode from the polymer under
the strong electric field. At around $t=1200$fs, the charge in the
polymer decreases to about one electron and a well-shaped polaron
locates at the position close to the right interface. It lasts for
a few thousands femtoseconds, depending on the strength of the
applied field, before the charge enters completely the right metal
electrode and the polymer gets back to the dimerized state.

The result for the electric fields $E_0\ge 8.0\times 10^4$V/cm
indicates that the electric field applied at the polymer chain
reduces effectively the potential barrier at the metal/polymer
interface. To show that, we consider the case of a smaller voltage
bias applied at the left metal electrode. For example, we take
$V_0=0.3$eV, it is found that the charge can be injected into the
polymer for the electric fields $\geq 1.7\times 10^5$V/cm. In
Fig.~6, we show the result of the calculation at the electric
field $E_0=2.0\times 10^5$V/cm and the voltage bias $V_0=0.3$eV.
The physical process is similar but the polaron lasts only a short
period since the charge enters the right metal electrode sooner
under a stronger electric field, which also indicates that the
potential barrier at the polymer/metal interface is effectively
reduced.

\begin{figure}
\includegraphics[angle=270,scale=0.32]{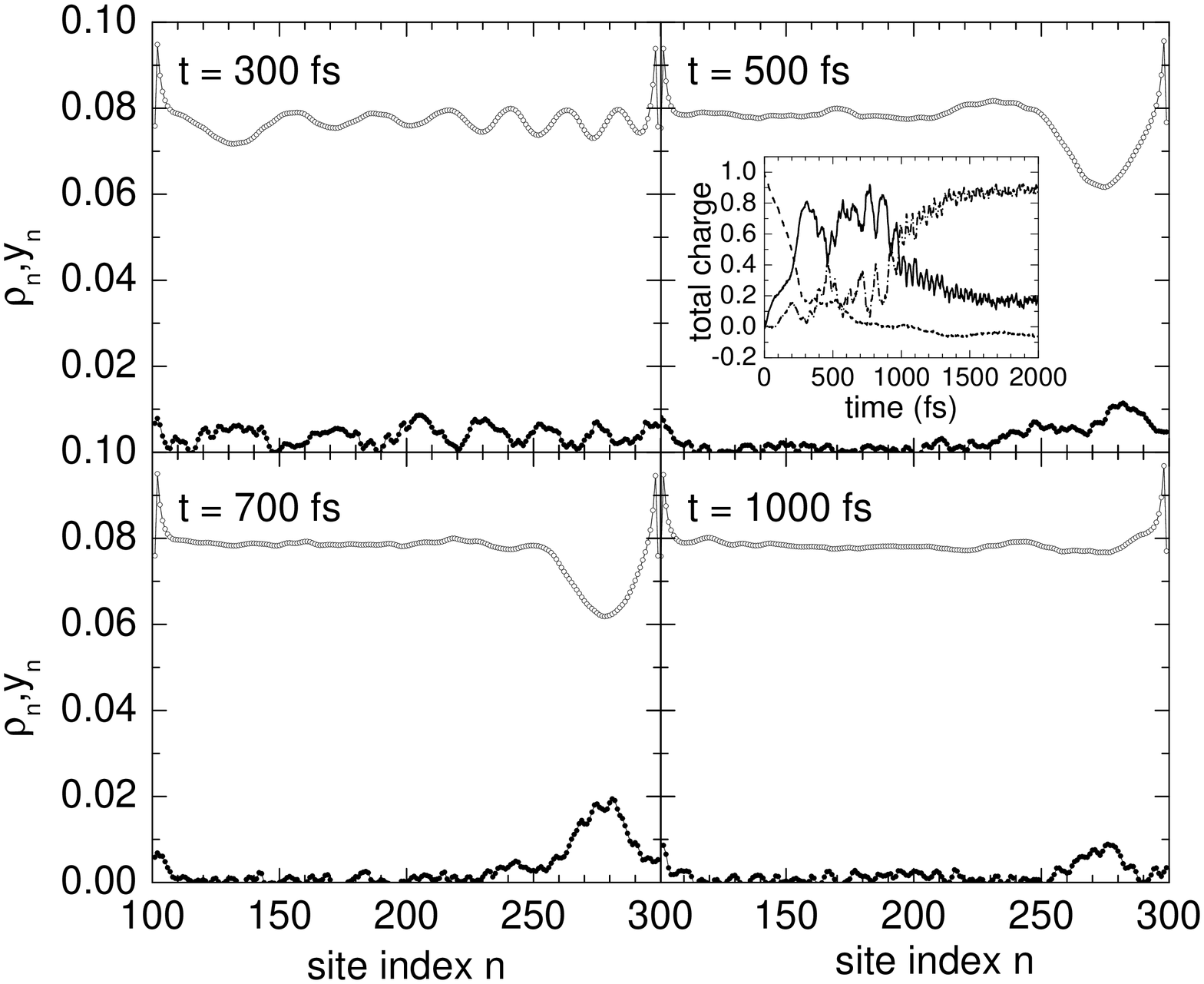}
\caption{Same as in Fig.~2 but with $E_0=2.0\times 10^5$V/cm and
$V_0=0.3$eV. The inset shows the total charges in the polymer
chain (solid line) and the metal electrodes (dash and dot-dash
lines). \label{fig6}}
\end{figure}

\section{summary}

In summary, we have investigated the polaron formation starting
from the charge injection by using a non-adiabatic dynamic method
based on the time-dependent Schr\"{o}dinger equation for the
electronic wavefunctions combining the Newtonian equation of
motion for the polymer monomer displacements. It is shown that the
polaron is formed while it moves slowly only under a weak electric
field $E_0<1.0\times 10^4$V/cm. But for a stronger electric field,
the polaron can be still formed because the charge is not easy to
be jumped away from the chain. This result implies that the low
mobility of the charge transport in polymers is mainly due to the
obstacle for carriers to travel through the polymer/metal
interface and/or jump over a neighboring chain. Our results also
show that the electric field applied at the polymer chain reduces
effectively the potential barrier in the metal/polymer interface.

\begin{acknowledgments}
This work was partially supported by National Natural Science
Foundation of China (No.~90103034) and the State Ministry of
Education of China (No.~20020246006). One of the authors (C.Q.W.)
is grateful to the Institute of Materials Structure Science of KEK
for the hospitality during his visit there.
\end{acknowledgments}

\end{document}